\documentclass{webofc}
\usepackage[varg]{txfonts}   
\usepackage{graphicx}
\def\bc{\begin{center}}
\def\ec{\end{center}}
\def\beq{\begin{equation}}
\def\eeq{\end{equation}}
\def\beq*{\begin{equation*}}
\def\eeq*{\end{equation*}}

\def\hs#1{\hspace*{#1cm}}
\def\av#1{\langle #1 \rangle}

\def\omN{{\omega_N}}

\def\bc{\begin{center}}
\def\ec{\end{center}}
\def\beq{\begin{equation}}
\def\eeq{\end{equation}}

\def\hs#1{\hspace*{#1cm}}
\def\av#1{\langle #1 \rangle}

\def\Deta{\Delta\eta}

\def\deta{\delta\eta}

\def\omN{{\omega_N}}

\begin{document}
%

\title{On the  Interpretation of the Balance Function}
%
%

\author{\firstname{Vladimir} \lastname{Vechernin}
\inst{1}
\fnsep\thanks{\email{v.vechernin@spbu.ru}}}

\institute{Saint Petersburg State University}

\abstract{%
We construct a simple toy model and explicitly demonstrate
that the Balance Function (BF) can become negative for some values of the rapidity separation 
and hence can not have any probabilistic interpretation.
In particular, the BF can not be interpreted as
the probability density for the balancing charges to occur separated by the given rapidity interval.
}
\maketitle
%

\section{Introduction} 
\label{sec-intro}

In experiments, the netcharge fluctuations usually are studied
\cite{Voloshin02,ALICE16} by calculating the quantity $\nu_{dyn}$, defined as:
\beq
\label{nu-dyn}
\nu_{dyn}(\deta)\equiv\frac{\av{n_+(n_+ -1)}}{\av{n_+}^2}+\frac{\av{n_-(n_- -1)}}{\av{n_-}^2}
-2\frac{\av{n_+n_-}}{\av{n_+}\av{n_-}} \ ,
\eeq
where
$n^+$ and $n^-$
is a number of positive and negative particles observed in the pseudorapidity interval  $\deta$.
In some cases there is more convenient to modify the normalization of this quantity
introducing \cite{Altsybeev19}:
\beq
\label{nuS-def}
\nu_s(\deta)\equiv-\,\frac{\av{n_+}+\av{n_-}}{4}
\nu_{dyn}(\deta) \ .
\eeq

This variable is closely connected with the so-called Balance Function (BF) \cite{Bass00},
usually defined as
\beq
\label{BF-def}
B(\eta_1, \eta_2)= \frac{1}{2}
\left[
  \frac{\rho_{+-} (\eta_1, \eta_2)}{\rho_{+} (\eta_1)}
+\frac{\rho_{-+} (\eta_1, \eta_2)}{\rho_{-} (\eta_1)}
-\frac{\rho_{++} (\eta_1, \eta_2)}{\rho_{+} (\eta_1)}
-\frac{\rho_{--} (\eta_1, \eta_2)}{\rho_{-} (\eta_1)}
 \right]  \ ,
\eeq
where $\rho_{+} (\eta)$, $\rho_{+-} (\eta_1, \eta_2)$ and so on are
the inclusive and double inclusive pseudorapidity distributions of corresponding charged particles
(on the correspondence with other possible alternative definitions of the BF see e.g. \cite{Univ19}).

In the most simple way the connection between the $\nu_s(\deta)$ and  the $B(\eta_1, \eta_2)$
can be established
in mid-rapidity region at LHC energies,
where the translation invariance in rapidity is valid. In this case
the single inclusive distributions are constant: $\rho_+(\eta)  = \av{n_+}/\deta$, $\rho_-(\eta)  = \av{n_-}/\deta$
and the double inclusive distributions depend only on the differences of their arguments:
$\rho_{+-} (\eta_1, \eta_2)=\rho_{+-} (\eta_1- \eta_2)$ and so on.
Hence the BF also will depend only
on the $\eta_1- \eta_2\equiv\Deta$.

The charge symmetry is also well satisfied in this case:
\beq
\label{sym-ch}
\av{n_+}=\av{n_-}  \ ,  \hs{1} \omega_{n_+}=\omega_{n_-}   \ ,  \hs{1} \omega_{n_+}\equiv D_{n_+}/\av{n_+}
 \ ,  \hs{0.5}  D_{n_+}\equiv \av{n^2_+} -\av{n_+}^2 \ .
\eeq
Then the expressions (\ref{nuS-def}) for  $\nu_s(\deta)$
and (\ref{BF-def}) for $B(\eta_1, \eta_2)$ reduce to
\beq
\label{nuS}
\nu_s(\deta)=\frac{\av{n_+n_-}-\av{n_+(n_+ -1)}}{\av{n_+}}
=1+\frac{\av{n_+n_-}-\av{n_+^2}}{\av{n_+}}
\eeq
and
\beq
\label{BF}
B(\eta_1-\eta_2)=
  \frac{\rho_{+-} (\eta_1- \eta_2)-\rho_{++} (\eta_1-\eta_2)}{\rho^0_{+}}
  \ .
\eeq
Then by the direct integration of (\ref{BF}) we get
\beq
\label{nuS-BF-gen}
\nu_s(\deta)=\frac{1}{\deta}
\int_{\deta}\!\!\!d\eta_1 \int_{\deta} \!\!\!d\eta_2
\ B(\eta_1-\eta_2)  \ ,
\eeq
where we have taken into account the normalization conditions (\ref{rho-pp-norm}) and (\ref{rho-pm-norm})
(see the next Section).

Since by the definition (\ref{BF-def}) the BF is symmetric: $B(\Deta)=B(-\Deta)$, the integral (\ref{nuS-BF-gen})
can be written as follows (see e.g. the Appendix A in the paper \cite{NPA15}):
$$
\nu_s(\deta)
=\frac{1}{\deta}
\int_{\eta}^{\eta+\deta}\!\!\!\!\!\!d\eta_1 \int_{\eta}^{\eta+\deta} \!\!\!\!\!\!d\eta_2
\ B(\eta_1-\eta_2)
=\frac{1}{\deta}
\int_{-\deta/2}^{\deta/2}\!\!\!d\eta_1 \int_{-\deta/2}^{\deta/2} \!\!\!d\eta_2
\ B(\eta_1-\eta_2)
$$
\beq
\label{nuS-BF-tr}
=\frac{1}{\deta}
\int_{-\deta}^{\deta} \!\!\!d(\Deta)
\ B(\Deta)\ t_{\deta}(\Deta)
=\frac{2}{\deta}
\int_{0}^{\deta} \!\!\!d(\Deta)
\ B(\Deta)\ (\deta-\Deta)
\ ,
\eeq
where the $t_{\deta}(\Deta)$ is the usual phase space triangular weight function:
\beq
\label{tri}
t_{\deta}(\Deta)= [\theta(-\Deta)(\deta+\Deta) +\theta(\Deta)(\deta-y)]\,\theta(\deta-|\Deta|) \geq 0 
\eeq
(see the Fig. A.1 in the paper \cite{NPA15}).

In paper \cite{Bass00} the authors state that
"The BF would represent the probability that the balancing charges were separated by $\Deta$
(in our formalism we include a division by $\Deta$ to express $B(\Deta)$ as a density)."
Nevertheless in the Introduction Section of the paper \cite{ALICE16} it is mentioned that
the value of $\nu_{dyn}(\deta)$ can be both negative and positive:
"A negative value of $\nu_{dyn}$ signifies the dominant contribution from correlations between
pairs of opposite charges. On the other hand,
a positive value indicates the significance of the same charge pair correlations."

By formula (\ref{nuS-def}) this means that in some cases the $\nu_s(\deta)$
can take negative values.
Then by formula (\ref{nuS-BF-tr}) we see that in this case
the BF must also be negative at least at some values of $\Deta$
to ensure the negative value of the integral (\ref{nuS-BF-tr}),
as the triangular weight function (\ref{tri}) is positive: $t_{\deta}(\Deta) \geq 0$.
But if the $\nu_s(\deta)$ and the BF $B(\Deta)$ can take negative values
they can not have any probabilistic interpretation,
in particular that mentioned in paper \cite{Bass00}.

In present short note we explicitly confirm this fact by direct calculations
for very simple toy model.

Note also in the conclusion of this section that the $\nu_s(\deta)$ is simply
connected with the variable $\Sigma(n_+,n_-)$,
\beq
\label{nuS-Sigma}
\nu_s(\deta)=1-\Sigma(n_+,n_-) \ ,
\eeq
denoted in \cite{EPJA19} as $\Sigma(n_{F}^{+},n_{F}^{-})$.

\section{Model independent definitions and relations} 
\label{sec-def}

We start with the definitions of inclusive and double inclusive pseudorapidity distributions of charged particles:
\begin{equation}
\label{rho12-def}
\rho_\pm (\eta)\equiv\frac{dN^\pm_{ch}}{d\eta}
\ , \hs1
\rho_{++} (\eta_1, \eta_2)\equiv\frac{d^2N^{++}_{ch}}{d\eta_1\,d\eta_2}
\ , \hs1
\rho_{+-} (\eta_1, \eta_2)\equiv\frac{d^2N^{+-}_{ch}}{d\eta_1\,d\eta_2}   \ ,
\end{equation}
which are normalized as follows:
\beq
\label{rho1-norm}
\int_{\deta}\!\!\!d\eta\  \rho_\pm (\eta)  = \av{n_\pm}  \ ,
\eeq
\beq
\label{rho-pp-norm}
\int_{\deta}\!\!\!d\eta_1 \int_{\deta} \!\!\!d\eta_2
\ \rho_{++} (\eta_1, \eta_2) = \av{n_+(n_+-1)} \ .
\eeq
\beq
\label{rho-pm-norm}
\int_{\deta}\!\!\!d\eta_1 \int_{\deta} \!\!\!d\eta_2
\ \rho_{+-} (\eta_1, \eta_2) = \av{n_+n_-} \ .
\eeq
Then we define the two-particle correlation functions by a standard way, \cite{Voloshin02}:
\begin{equation}
\label{C}
C_{++}(\eta_1,\eta_2)\equiv
\frac{\rho_{++} (\eta_1, \eta_2)}{\rho_{+}(\eta_1) \rho_{+}(\eta_2)}-1
\ , \hs1
C_{+-}(\eta_1,\eta_2)\equiv
\frac{\rho_{+-} (\eta_1, \eta_2)}{\rho_{+}(\eta_1) \rho_{-}(\eta_2)}-1
\ .
\end{equation}

In mid-rapidity region at LHC energies, when the translation invariance in rapidity
and the charge symmetry, mentioned above, take place, these formulae can be simplified,
using that
\begin{equation}
\label{rho-tr-inv}
\rho_+(\eta)  =\rho_- (\eta)  = \rho^0_+  = const= \av{n_+}/\deta
\ ,
\end{equation}
$$
\rho_{++} (\eta_1, \eta_2)=\rho_{++} (\eta_1- \eta_2)
\ , \hs1
\rho_{+-} (\eta_1, \eta_2)=\rho_{+-} (\eta_1- \eta_2)
$$
and hence
\begin{equation}
\label{C-tr-inv}
C_{++}(\eta_1,\eta_2)= C_{++}(\eta_1-\eta_2)
\ , \hs1
C_{+-}(\eta_1,\eta_2)= C_{+-}(\eta_1-\eta_2)
 \ .
\end{equation}

Then by (\ref{rho1-norm})-(\ref{C-tr-inv}) we have
\beq
\label{C-pp-norm}
\rho^0_+\rho^0_+\int_{\deta}\!\!\!d\eta_1 \int_{\deta} \!\!\!d\eta_2
\ C_{++} (\eta_1- \eta_2) = \av{n_+(n_+-1)}- \av{n_+}^2 \ .
\eeq
\beq
\label{C-pm-norm}
\rho^0_+\rho^0_-\int_{\deta}\!\!\!d\eta_1 \int_{\deta} \!\!\!d\eta_2
\ C_{+-} (\eta_1- \eta_2) = \av{n_+n_-}-\av{n_+}\av{n_-} \ .
\eeq
Using now definition (\ref{nuS}) we express the $\nu_s(\deta)$ through the
correlation functions $C_{+-}$ and $C_{++}$ by \emph{the model independent way}:
\beq
\label{nuS-C}
\nu_s(\deta)=\frac{\rho^0_+}{\deta}
\int_{\deta}\!\!\!d\eta_1 \int_{\deta} \!\!\!d\eta_2
\ \left[C_{+-} (\eta_1- \eta_2)- C_{++} (\eta_1- \eta_2) \right]  \ .
\eeq
Simultaneously from formula (\ref{BF}) for the BF we have
\beq
\label{BF-C}
B(\eta_1- \eta_2)=\rho^0_+ \cdot
\ \left[C_{+-} (\eta_1- \eta_2)- C_{++} (\eta_1- \eta_2) \right]  \ .
\eeq

\section{The models with independent identical sources} 
\label{sec-independ}
In models with independent identical sources the following formula \cite{NPA15}
for $C(\eta_1,\eta_2)$ takes place (see a simple proof in Appendix A):
\beq
\label{C-Lam}
C(\eta_1,\eta_2)=\frac{\Lambda(\eta_1,\eta_2)+\omN}{\av N}  \ ,
\eeq
where $N$ is a number of sources, which fluctuates event by event
around some mean value,  $\av N$, with some scaled variance, $\omN={D_N}/{\av{N}}$.

The $\Lambda(\eta_1,\eta_2)$ is the two-particle correlation function
characterizing a single source.
It is defined similarly to $C(\eta_1,\eta_2)$, but taking into account
only particles produced by a given source:
\begin{equation}
\label{Lam}
\Lambda_{++}(\eta_1,\eta_2)\equiv
\frac{\lambda_{++} (\eta_1, \eta_2)}{\lambda_{+}(\eta_1) \lambda_{+}(\eta_2)}-1
\ , \hs1
\Lambda_{+-}(\eta_1,\eta_2)\equiv
\frac{\lambda_{+-} (\eta_1, \eta_2)}{\lambda_{+}(\eta_1) \lambda_{-}(\eta_2)}-1
\ ,
\end{equation}
where
\begin{equation}
\label{lam12-def}
\lambda_\pm (\eta)\equiv\frac{dN^\pm_{ch}}{d\eta}
\ , \hs1
\lambda_{++} (\eta_1, \eta_2)\equiv\frac{d^2N^{++}_{ch}}{d\eta_1\,d\eta_2}
\ , \hs1
\lambda_{+-} (\eta_1, \eta_2)\equiv\frac{d^2N^{+-}_{ch}}{d\eta_1\,d\eta_2}   \ ,
\end{equation}
are inclusive and double inclusive pseudorapidity distributions of charged particles
\emph{produced by a given source}.
They are normalized as follows:
\beq
\label{lam1-norm}
\int_{\deta}\!\!\!d\eta\  \lambda_\pm (\eta)  = \av{\mu_\pm}  \ ,
\eeq
\beq
\label{lam-pp-norm}
\int_{\deta}\!\!\!d\eta_1 \int_{\deta} \!\!\!d\eta_2
\ \lambda_{++} (\eta_1, \eta_2) = \av{\mu_+(\mu_+-1)} \ .
\eeq
\beq
\label{lam-pm-norm}
\int_{\deta}\!\!\!d\eta_1 \int_{\deta} \!\!\!d\eta_2
\ \lambda_{+-} (\eta_1, \eta_2) = \av{\mu_+\mu_-} \ .
\eeq

In mid-rapidity region at LHC energies, when the translation invariance in rapidity
and the charge symmetry take place, these formulae can again be simplified,
using that
\begin{equation}
\label{lam-tr-inv}
\lambda_+(\eta)  =\lambda_- (\eta)  = \lambda^0_+ = \lambda^0_-  = const
= \frac{\av{\mu_+}}{\deta}= \frac{\av{n_+}}{\deta \av{N}}=\frac{\rho^0_+}{\av{N}}
\ ,
\end{equation}
$$
\lambda_{++} (\eta_1, \eta_2)=\lambda_{++} (\eta_1- \eta_2)
\ , \hs1
\lambda_{+-} (\eta_1, \eta_2)=\lambda_{+-} (\eta_1- \eta_2)
$$
and hence
\begin{equation}
\label{Lam-tr-inv}
\Lambda_{++}(\eta_1,\eta_2)= \Lambda_{++}(\eta_1-\eta_2)
\ , \hs1
\Lambda_{+-}(\eta_1,\eta_2)= \Lambda_{+-}(\eta_1-\eta_2)
 \ .
\end{equation}
Then
\begin{equation}
\label{Lam-lam}
\Lambda_{++}(\eta_1-\eta_2)=
\frac{\lambda_{++} (\eta_1-\eta_2)}{\lambda^0_+\lambda^0_+}-1
\ , \hs1
\Lambda_{+-}(\eta_1-\eta_2)=
\frac{\lambda_{+-} (\eta_1- \eta_2)}{\lambda^0_+\lambda^0_-}-1 \ .
\end{equation}

Substituting now the general connection (\ref{C-Lam})
into formula  (\ref{nuS-C}) we finally express the $\nu_s(\deta)$ through the
correlation functions $\Lambda_{+-}$ and $\Lambda_{++}$ of a single source:
\beq
\label{nuS-Lam}
\nu_s(\deta)=\frac{\lambda^0_+}{\deta}
\int_{\deta}\!\!\!d\eta_1 \int_{\deta} \!\!\!d\eta_2
\ \left[\Lambda_{+-} (\eta_1- \eta_2)- \Lambda_{++} (\eta_1- \eta_2) \right]  \ .
\eeq
Note that a dependence on  $\av N$ and $\omN={D_N}/{\av{N}}$
is canceled what  proves the strongly intensive behavior of this variable
in the case with identical sources.

We see this also from the fact that formula (\ref{nuS-Lam})
coincides with the definition (\ref{nuS}) when replacing all engaged quantities
by the corresponding ones for one source.
That also can be written as
\beq
\label{nuS-m-mu}
\nu_s(\deta)=\frac{\av{n_+n_-}-\av{n_+(n_+ -1)}}{\av{n_+}}
=\frac{\av{\mu_+\mu_-}-\av{\mu_+(\mu_+ -1)}}{\av{\mu_+}}
\eeq
in any model with identical courses.

As mentioned in the Introduction the $\nu_s(\deta)$ is simply connected
with the balance function $B(\eta_1-\eta_2)$.
In any model with the identical independent sources
in the central region, where
the translation invariance in rapidity and the charge symmetry take place,
we have (see e.g. Section 5 of the paper \cite{Univ19}):
\beq
\label{BF-Lam}
B(\eta_1-\eta_2)= \lambda^0_+ \cdot
\ \left[\Lambda_{+-} (\eta_1- \eta_2)- \Lambda_{++} (\eta_1- \eta_2) \right] \ .
\eeq
One can immediately obtain this formula substituting (\ref{C-Lam}) into (\ref{BF-C})
and taking into account the relation (\ref{lam-tr-inv}).

Comparing formulas (\ref{nuS-Lam}) and (\ref{BF-Lam}) we see
that the general relation (\ref{nuS-BF-gen}):
$$
\nu_s(\deta)=\frac{1}{\deta}
\int_{\deta}\!\!\!d\eta_1 \int_{\deta} \!\!\!d\eta_2
\ B(\eta_1-\eta_2)  \ ,
$$
of course, is true in this particular case.

\section{Toy model with production of correlated charge pairs by a source} 
\label{sec-toy-uncorr}
Let us consider at first the very simple model,
when each source always produces only one plus-minus pair,
with plus and minus particles being uniformly distributed
in some wide interval $(-Y/2, Y/2)$, $Y\gg 1$.

In this simple model
\begin{equation}
\label{mod1-1}
\lambda^0_+=\frac{1}{Y}
\ , \hs1
\lambda_{++} (\eta_1- \eta_2)=0
\ , \hs1
\lambda_{+-} (\eta_1- \eta_2)=\frac{1}{Y^2}
\ .
\end{equation}
To test these formulae we can use the normalization conditions
(\ref{lam1-norm})-(\ref{lam-pm-norm}) in the whole acceptance $Y$:
\beq
\label{lam1-Y}
\int_{Y}\!\!\!d\eta\  \lambda_\pm (\eta)  = 1  \ ,
\eeq
\beq
\label{lam-pp-Y}
\int_{Y}\!\!\!d\eta_1 \int_{Y} \!\!\!d\eta_2
\ \lambda_{++} (\eta_1- \eta_2) = 0 \ .
\eeq
\beq
\label{lam-pm-Y}
\int_{Y}\!\!\!d\eta_1 \int_{Y} \!\!\!d\eta_2
\ \lambda_{+-} (\eta_1- \eta_2) = 1 \ .
\eeq

Then by (\ref{Lam-lam}) we have
\begin{equation}
\label{mod1-2}
\Lambda_{++}(\eta_1-\eta_2)=-1
\ , \hs1
\Lambda_{+-}(\eta_1-\eta_2)=0
\ .
\end{equation}
As expected we see no correlation between plus and minus particles
produced from the same source, $\Lambda_{+-}(\eta_1-\eta_2)=0$,
and a strong anticorrelation between plus and plus particles
from one source,  $\Lambda_{++}(\eta_1-\eta_2)=-1$, because
the only one plus particle, produced from a source, can't be simultaneously
at both $\eta_1$ and $\eta_2$ pseudorapidities.

Substituting all this now in formula (\ref{nuS-Lam}) we find
\beq
\label{mod1-3}
\nu_s(\deta)=\frac{1}{Y\deta}\, \deta^2 \left[0-(-1) \right]=\frac{\deta}{Y}  \ .
\eeq
The interpretation of the $\nu_s(\deta)=\frac{\deta}{Y}$ as the probability
to find the negatively charged particle in the rapidity interval  $\deta$
under condition that we have already the positively charged particle in this interval
looks very suspicious. Since,  as we can see from formulae (\ref{mod1-2}) and (\ref{mod1-3}),
this result arises not due to correlation between plus and minus particles
but due to a strong  anticorrelation between plus and plus particles
in this simple model.

To verify these suspicions
let us consider a little bit more sophisticated model,
when each source always produces two plus-minus pairs,
with two plus and two minus particles being uniformly distributed
in some wide interval $(-Y/2, Y/2)$, $Y\gg 1$.

In this version of the model
\begin{equation}
\label{mod2-1}
\lambda^0_+=\frac{2}{Y}
\ , \hs1
\lambda_{++} (\eta_1- \eta_2)=\frac{2}{Y^2}
\ , \hs1
\lambda_{+-} (\eta_1- \eta_2)=\frac{4}{Y^2}
\ .
\end{equation}
Again we can test these formulae using the normalization conditions
(\ref{lam1-norm})-(\ref{lam-pm-norm}) in the whole acceptance $Y$:
\beq
\label{lam1-Y2}
\int_{Y}\!\!\!d\eta\  \lambda_\pm (\eta)  = 2  \ ,
\eeq
\beq
\label{lam-pp-Y2}
\int_{Y}\!\!\!d\eta_1 \int_{Y} \!\!\!d\eta_2
\ \lambda_{++} (\eta_1-\eta_2) = 2 \ .
\eeq
\beq
\label{lam-pm-Y2}
\int_{Y}\!\!\!d\eta_1 \int_{Y} \!\!\!d\eta_2
\ \lambda_{+-} (\eta_1-\eta_2) = 4 \ .
\eeq

Then by (\ref{Lam-lam}) we have
\begin{equation}
\label{mod2-2}
\Lambda_{++}(\eta_1-\eta_2)=-\frac{1}{2}
\ , \hs1
\Lambda_{+-}(\eta_1-\eta_2)=0
\ .
\end{equation}
As expected again we see no correlation between plus and minus particles
produced from the same source, $\Lambda_{+-}(\eta_1-\eta_2)=0$,
and attenuation of the anticorrelation between plus and plus particles
from one source,  $\Lambda_{++}(\eta_1-\eta_2)=-\frac{1}{2}$, because
now two plus particles are produced from a source and
$\lambda_{++} (\eta_1- \eta_2)=\frac{2}{Y^2}>0$.

Substituting all this in formula (\ref{nuS-Lam}) we find that again
\beq
\label{mod2-3}
\nu_s(\deta)=\frac{2}{Y\deta}\, \deta^2 \left[0-\left(-\frac{1}{2}\right) \right]=\frac{\deta}{Y}  \ .
\eeq

It is easy to prove that
in model,
when each source always produces $k$ plus-minus pairs,
with $k$ plus and $k$ minus particles being uniformly distributed
in some wide interval $(-Y/2, Y/2)$, $Y\gg 1$, we'll have
\beq
\label{modk-3}
\nu_s(\deta)=\frac{k}{Y\deta}\, \deta^2 \left[0-\left(-\frac{1}{k}\right) \right]=\frac{\deta}{Y}  \ .
\eeq

The interpretation of the $\nu_s(\deta)=\frac{\deta}{Y}$ as the probability
to find the negatively charged particle in the rapidity interval  $\deta$
under condition that we have already the positively charged particle in this interval
still holds, since in each event we have equal number of plus and minus particles
uniformly distributed
in some wide interval $(-Y/2, Y/2)$, $Y\gg 1$,  as in the initial version of the model 
with one charge pair production by a source.
Nevertheless it looks strange since
it based not on correlations between plus and minus particles
but on anticorrelations between plus and plus particles in this simple model.

Note that in this case by formula (\ref{BF-Lam}) the BF ("the probability density") is equal to $1/Y$:
\beq
\label{BF-indep-pairs}
B(\Deta)
=\frac{k}{Y}\,  \left[0-\left(-\frac{1}{k}\right) \right]=\frac{1}{Y} \ ,
\eeq
what after integration over rapidity interval $\deta$ by (\ref{nuS-BF-tr})
again leads to the formula (\ref{modk-3}).

\section{Toy model with production of correlated charge pairs by a source} 
\label{sec-toy-corr}

As we can see in previous section
this result, $\nu_s(\deta)=\frac{\deta}{Y}$, arises due to
plus-plus anticorrelation,
$\Lambda_{++}(\eta_1-\eta_2)=-\frac{1}{k}$,
in the version of the model with production of $k$ independent
plus-minus pairs by each source.
After multiplying by $\lambda^0_+=\frac{k}{Y}$
and the integration we just have this result.

So, in present section we are trying to introduce some additional plus-plus correlation,
formulating a more complex version of the model.

\subsection{Strong correlation between identical charges from a source} 
\label{subsec-toy-corr-strong}

Let us consider at first the model in which each source always produces two plus-minus pairs,
so that the rapidities of both  positive particles coincide
and the same is true for both minus particles 
(the maximally strong correlation between identical charges),
whereas the rapidities of the plus pair and the minus pair
themselves are uniformly distributed
in some wide interval $(-Y/2, Y/2)$, $Y\gg 1$.

In this version of the model
\begin{equation}
\label{mod3-1}
\lambda^0_+=\frac{2}{Y}
\ , \hs1
\lambda_{++} (\eta_1- \eta_2)=\frac{2}{Y}\delta(\eta_1- \eta_2)
\ , \hs1
\lambda_{+-} (\eta_1- \eta_2)=\frac{4}{Y^2}
\ .
\end{equation}
Again we can test these formulae using the normalization conditions
(\ref{lam1-norm})-(\ref{lam-pm-norm}) in the whole acceptance $Y$:
\beq
\label{lam-Y3}
\int_{Y}\!\!\!d\eta\  \lambda_\pm (\eta)  = 2  \ ,
\eeq
\beq
\label{lam-pp-Y3}
\int_{Y}\!\!\!d\eta_1 \int_{Y} \!\!\!d\eta_2
\ \lambda_{++} (\eta_1-\eta_2) = 2 \ .
\eeq
\beq
\label{lam-pm-Y3}
\int_{Y}\!\!\!d\eta_1 \int_{Y} \!\!\!d\eta_2
\ \lambda_{+-} (\eta_1-\eta_2) = 4 \ .
\eeq

Then by (\ref{Lam-lam}) we have
\begin{equation}
\label{mod3-2}
\Lambda_{++}(\eta_1-\eta_2)=\frac{Y}{2}\delta(\eta_1- \eta_2)-1
\ , \hs1
\Lambda_{+-}(\eta_1-\eta_2)=0
\ .
\end{equation}
As expected we see again no correlation between plus and minus particles
produced from the same source,
but we see now strong additional $\frac{Y}{2}\delta(\eta_1- \eta_2)$
correlation between positive particles from one source.

Substituting all this in formula (\ref{nuS-Lam}) we find
\beq
\label{nuS-3}
\nu_s(\deta)=\frac{2}{Y\deta} \left[0\cdot\deta^2-\left(\frac{Y}{2}\cdot \deta-1\cdot \deta^2\right) \right]
=\frac{2\deta}{Y}-1  \ .
\eeq
Before to make any conclusions we verify this important result
using the simple formula  (\ref{nuS-m-mu}):
$$
\nu_s(\deta)=\frac{\av{n_+n_-}-\av{n_+(n_+ -1)}}{\av{n_+}}
=\frac{\av{\mu_+\mu_-}-\av{\mu_+(\mu_+ -1)}}{\av{\mu_+}}
$$

In this version of the model
\beq
\label{mu-3}
\av{\mu_+}=\sum_{\mu_+\geq 1} P(\mu_+) \mu_+
=P(1) \cdot 1 + P(2) \cdot 2
=0 \cdot 1 + \frac{\deta}{Y} \cdot 2
= 2\frac{\deta}{Y}
\ ,
\eeq
\beq
\label{mu-pm-3}
\av{\mu_+\mu_-}=\sum_{\mu_+\geq 1;\mu_-\geq 1} P(\mu_+,\mu_-) \mu_+,\mu_-
=P(1,1) \cdot 1 + P(1,2) \cdot 2+ P(2,1) \cdot 2+ P(2,2) \cdot 4
\eeq
$$
=0 \cdot 1 + 0 \cdot 2+ 0 \cdot 2+ \frac{\deta}{Y}\frac{\deta}{Y} \cdot 4
= 4\left(\frac{\deta}{Y}\right)^2
\ ,
$$
\beq
\label{mu-pp-3}
\av{\mu_+(\mu_+-1)}=\sum_{\mu_+\geq 2} P(\mu_+) \mu_+(\mu_+-1)
=P(2) \cdot 2
=\frac{\deta}{Y} \cdot 2
= 2\frac{\deta}{Y}
\ .
\eeq
Then by formula  (\ref{nuS-m-mu}) we find
\beq
\label{nuS-3a}
\nu_s(\deta)
=\frac{\av{\mu_+\mu_-}-\av{\mu_+(\mu_+ -1)}}{\av{\mu_+}}
=\frac{4\left(\frac{\deta}{Y}\right)^2-2\frac{\deta}{Y}}{2\frac{\deta}{Y}}
=\frac{2\deta}{Y}-1  \ ,
\eeq
that coincides with (\ref{nuS-3}).

So, concluding we see that although
for the whole interval at $\deta=Y$ we have $\nu_s(Y)=1$,
as expected,
nevertheless the value of  the $\nu_s(\deta)$ at $\deta<Y/2$
becomes negative and hence can not have any probabilistic
interpretation.

Note that by formula (\ref{BF-Lam}) the BF in this case is as follows
\beq
\label{BF-corr-pairs}
B(\Deta)
=-\delta(\Deta)+\frac{2}{Y}
 \ ,
\eeq
what after integration over rapidity interval $\deta$ by (\ref{nuS-BF-tr})
again leads to the formula (\ref{nuS-3a}).

\subsection{Gentle correlation between identical charges from a source} 
\label{subsec-toy-corr-gentl}

From the model construction it is clear that if we'll use instead of the $\delta$-function
any enough narrow distribution normalized by unity, we'll arrive to the same conclusion.
Really, let us use in this subsection instead of the $\delta$-function
the step distribution normalized to unity and
spread over interval from $-a$ to $a$ ($a>0$):
\beq
\label{step}
\delta(\Deta)\ \to \ h_a(\Deta)\equiv\frac{1}{2a}\theta(a-|\Deta|) \ ,
\eeq

In this case for the version of the model, described in the previous Subsection \ref{subsec-toy-corr-strong},
we have
\begin{equation}
\label{mod3-1a}
\lambda^0_+=\frac{2}{Y}
\ , \hs1
\lambda_{++} (\eta_1- \eta_2)=\frac{2}{Y-a/2}\ h_a(\eta_1- \eta_2)
\ , \hs1
\lambda_{+-} (\eta_1- \eta_2)=\frac{4}{Y^2}
\ .
\end{equation}
Using the formulae (\ref{nuS-BF-tr}) we can check that the factor $2/(Y-a/2)$ ensures
the correct normalization condition (\ref{lam-pp-norm}) for the $\lambda_{++}(\eta_1- \eta_2)$:
\beq
\label{lam-Y3a}
\int_{Y}\!\!\!d\eta\  \lambda_\pm (\eta)  = 2  \ ,
\eeq
\beq
\label{lam-pp-Y3a}
\int_{Y}\!\!\!d\eta_1 \int_{Y} \!\!\!d\eta_2
\ \lambda_{++} (\eta_1-\eta_2)=
\eeq
$$
=\int_{-Y/2}^{Y/2}\!\!\!d\eta_1 \int_{-Y/2}^{Y/2} \!\!\!d\eta_2
\ \lambda_{++} (\eta_1-\eta_2)
=\frac{2}{Y-a/2} \int_{-Y}^Y  \!\!\!d(\Deta)
\ h_a(\Deta)
\ t_Y(\Deta)
= 2 \ ,
$$
\beq
\label{lam-pm-Y3a}
\int_{Y}\!\!\!d\eta_1 \int_{Y} \!\!\!d\eta_2
\ \lambda_{+-} (\eta_1-\eta_2) = 4 \ .
\eeq
Then by (\ref{Lam-lam}) we have
\begin{equation}
\label{mod3-2a}
\Lambda_{++}(\eta_1-\eta_2)=\frac{Y^2}{2Y-a}h_a(\eta_1- \eta_2)-1
\ , \hs1
\Lambda_{+-}(\eta_1-\eta_2)=0
\ .
\end{equation}

By the formula (\ref{BF-Lam}) we find now  the BF:
\beq
\label{BF-corr-pairs-a}
B(\Deta)
=-\ \frac{Y}{Y-a/2}\ h_a(\Deta)+\frac{2}{Y}
 \ .
\eeq
For $|\Deta|<a$ by (\ref{step}) we have
\beq
\label{BF-corr-pairs-a1}
B(\Deta)
=-\ \frac{Y}{(2Y-a)\,a}+\frac{2}{Y} \ .
\eeq
It is easy to check that at $|\Deta|<a<(1-1/\sqrt{2})\,Y\approx 0.29\,Y$ the BF is negative, $B(\Deta)<0$,
and can't be interpreted as a probability density.

We can now calculate $\nu_s(\deta)$ by the integration of the expression (\ref{BF-corr-pairs-a})
over rapidity interval $\deta$ using the formulae (\ref{nuS-BF-gen}) and (\ref{nuS-BF-tr}):
\beq
\label{nuS-corr-pairs-a}
\nu_s(\deta)
=\frac{1}{\deta} \int_{-\deta}^{\deta}  \!\!\!d(\Deta)
\ B(\Deta)
\ t_{\deta}(\Deta)=
\eeq
$$
=\frac{2\deta}{Y}
 - \frac{Y}{(2Y-a)\,a\,\deta} \int_{-\deta}^{\deta}  \!\!\!d(\Deta)
\ \theta(a-|\Deta|)
\ t_{\deta}(\Deta) \ .
$$
Then we find
\beq
\label{nuS-corr-pairs-a1}
\nu_s(\deta)
=\frac{2\deta}{Y}-\frac{2-a/\deta}{2-a/Y}
\hs 2 \textrm{at}\ \ \  \deta>a
\ ,
\eeq
and
\beq
\label{nuS-corr-pairs-a2}
\nu_s(\deta)
=\frac{2\deta}{Y}-\frac{\deta/a}{2-a/Y}
=\left( \frac{2}{Y}-\frac{Y}{(2Y-a)\,a} \right)\deta
 \hs 1 \textrm{at}\ \ \  \deta<a
\ .
\eeq
From formula (\ref{nuS-corr-pairs-a1}) we see that at $a\to 0$ the $\nu_s(\deta)$
go to the result (\ref{nuS-3a}), obtained in previous Subsection \ref{subsec-toy-corr-strong}.

By formula  (\ref{nuS-corr-pairs-a2}) we see that at
\beq
\label{cond}
\deta<a<(1-1/\sqrt{2})\,Y\approx 0.29\,Y
\eeq
(the same condition as the condition obtained from formula (\ref{BF-corr-pairs-a1}) for the BF)
the $\nu_s(\deta)$ is negative, $\nu_s(\deta)<0$,
and can't be interpreted as a probability.
Note that occurs for rather wide correlation function
$\lambda_{++} (\eta_1- \eta_2)$ (\ref{mod3-1a}),
with $a$ compared to $Y$, as follows from condition (\ref{cond}).

In conclusion of this subsection we perform one more check
of the obtained  formulae. Clear that this model with $a=Y$ corresponds
to the absence of the correlation between the same charge particles
from a source. Hence, in this case we have a source always emitting
 two pairs of uncorrelated plus-minus particles. This version of the model
 was already considered in Section \ref{sec-toy-uncorr} (the case with $k=2$).

Really, if in formula (\ref{mod3-1a}) we put $a=Y$,
then the formula for $\lambda_{++} (\eta_1- \eta_2)$, (\ref{mod3-1a}),
reduces to (\ref{mod2-1}):
$$
\lambda_{++} (\eta_1- \eta_2)=\frac{2}{Y-a/2}\ h_a(\eta_1- \eta_2)
\ \to\
\frac{2}{Y^2} \ .
$$
The formula for $\Lambda_{++}(\eta_1-\eta_2)$, (\ref{mod3-2a}),
reduces to (\ref{mod2-2}):
$$
\Lambda_{++}(\eta_1-\eta_2)=\frac{Y^2}{2Y-a}h_a(\eta_1- \eta_2)-1
\ \to\
-\ \frac{1}{2} \ .
$$
The formula for $B(\Deta)$, (\ref{BF-corr-pairs-a}),
reduces to (\ref{BF-indep-pairs}):
$$
B(\Deta)=-\ \frac{Y}{Y-a/2}\ h_a(\Deta)+\frac{2}{Y}
\ \to\
\frac{1}{Y} \ .
$$
The formula for $\nu_s(\deta)$, (\ref{nuS-corr-pairs-a2}),
reduces to (\ref{mod2-3}):
$$
\nu_s(\deta)
=\left( \frac{2}{Y}-\frac{Y}{(2Y-a)\,a} \right)\deta
\ \to\
\frac{\deta}{Y} \ .
$$

So, we see that the model considered in this Subsection \ref{subsec-toy-corr-gentl}
with gentle correlation between identical charges,
given by the function (see formulas (\ref{step}) and (\ref{mod3-1a})):
\beq
\label{lam-step}
\lambda_{++} (\Deta)=\frac{2}{Y-a/2}\ h_a(\Deta)=\frac{1}{(2Y-a)\,a}\ \theta(a-|\Deta|) \ ,
\eeq
with arbitrary value of the correlation width parameter $a$, $0<a\leq Y$,
on the one hand at $a\to 0$ go to the model
with strong correlation between identical charges
 considered in the previous Subsection \ref{subsec-toy-corr-strong},
and on the other hand at $a=Y$ go to the model
with uncorrelated charge pairs production by a source
considered in the Section \ref{sec-toy-uncorr}.

We see also that the negative values of the BF $B(\Deta)$ at $\Deta<a$
and the $\nu_s(\deta)$ at $\deta<a$ already occur 
when we introduce the rather weak correlation between
same charge particles
with the value of $a$ compared to $Y$, namely at $a<(1-1/\sqrt{2})\,Y\approx 0.29\,Y$
as follows from condition (\ref{cond}).

\section{Conclusion} 
\label{sec-concl}
In this short note by constructing of a simple toy model
we explicitly demonstrate that
the values of the $\nu_s(\deta)$ and hence the $\nu_{dyn}(\deta)$,
$$
\nu_s(\deta)\equiv-\,\frac{\av{n}}{4}
\nu_{dyn}(\deta) \ ,
$$
can be both negative and positive,
so it can not have any probabilistic interpretation,
as e.g. the probability
that the balancing charges occur in the same rapidity interval $\deta$.

By relation
$$
\nu_s(\deta)=\frac{1}{\deta}
\int_{\deta}\!\!\!d\eta_1 \int_{\deta} \!\!\!d\eta_2
\ B(\eta_1-\eta_2)  \ ,
$$
it follows 
that in this case
the BF must also be negative at least for some values of $\Deta=\eta_1-\eta_2$
to ensure the negative value of the integral. 
We also check it explicitly calculating the BF in our toy model.
 
But if the BF $B(\Deta)$ can take negative values
it can not have any probabilistic interpretation in general case.
In particular, the BF can not be interpreted as
the probability density for the balancing charges to occur separated by the rapidity
interval $\Deta$, as it was formulated in the paper \cite{Bass00}.

\section*{Appendix. A proof of the formula (\ref{C-Lam})} 
\label{sec-interpret}

For a class of events with fixed number of sources $N$, following the paper \cite{Voloshin02},  we have
\begin{equation}
\label{rhoN-a}
\rho^{(N)}(\eta)=N \lambda(\eta) \ ,
\end{equation}
\begin{equation}
\label{rho2N-a}
\rho_2^{(N)} (\eta_1, \eta_2)=N \lambda_2 (\eta_1, \eta_2)+ N(N-1) \lambda(\eta_1) \lambda(\eta_2) \ .
\end{equation}
Then averaging over events with different number of sources $N$ we find
\begin{equation}
\label{rho-a}
\rho(\eta)=\sum_N P(N) \rho^{(N)}(\eta)=\sum_N P(N) N \lambda(\eta) =\av N \lambda(\eta)\ ,
\end{equation}
\begin{equation}
\label{rho2-a}
\rho_2 (\eta_1, \eta_2)=\sum_N P(N) \rho_2^{(N)} (\eta_1, \eta_2)=\av N \lambda_2 (\eta_1, \eta_2)+ \av {N(N-1)} \lambda(\eta_1) \lambda(\eta_2) \ .
\end{equation}
Using now definitions (\ref{C}) and (\ref{Lam}) we have
\begin{equation}
\label{C-Lam-a}
C (\eta_1, \eta_2)=\frac{\rho_2 (\eta_1, \eta_2)}{\rho(\eta_1)\rho(\eta_2)}-1
=\frac{\av N [\lambda_2 (\eta_1, \eta_2)-\lambda(\eta_1) \lambda(\eta_2)]}
{\av N \lambda(\eta_1) \av N \lambda(\eta_2)} + \frac{\av {N^2}}{\av {N}^2}-1=
\end{equation}
$$
=\frac{\Lambda(\eta_1,\eta_2)}{\av N}+\frac{\av {N^2}-\av {N}^2}{\av {N}^2}
=\frac{\Lambda(\eta_1,\eta_2)+\omN}{\av N} \ .
$$


\section*{Acknowledgments} 
The author is grateful to Igor Altsybeev for stimulating discussions.
The research was supported by the Russian Foundation for Basic Research grant (No. 18-02-40075)
and the St. Petersburg State University grant for outgoing academic activity (Id: 41159705).
%
%

\end{document}